\documentclass[traditabstract]{aa} % for the abstract without structuration
\usepackage{graphicx}
\usepackage{natbib}
\usepackage{savesym}
\usepackage{txfonts}
\savesymbol{iint}
\savesymbol{iiint}
\savesymbol{iiiint}
\savesymbol{idotsint}
\usepackage{amsmath}
\restoresymbol{AMS}{iint}
\restoresymbol{AMS}{iiint}
\restoresymbol{AMS}{iiiint}
\restoresymbol{AMS}{idotsint}

\usepackage{bbm}
\usepackage{aas_macros}
\usepackage{dsfont}
\usepackage{fixltx2e}
\begin{document}
\title{Efficient least-squares basket-weaving}
\author{B. Winkel\inst{1}
          \and
         L. Fl\"{o}er\inst{2}
          \and
         A. Kraus\inst{1}
       }

\institute{Max-Planck-Institut f\"{u}r Radioastronomie (MPIfR), 
              Auf dem H\"{u}gel\,69, 53121 Bonn, Germany
              \and
              Argelander-Institut f\"{u}r Astronomie (AIfA), 
              Auf dem H\"{u}gel\,71, 53121 Bonn, Germany\\
              \email{bwinkel@mpifr.de}
}

\date{Preprint online version}

\abstract{We report on a novel method to solve the basket-weaving problem. Basket-weaving is a technique that is used to remove scan-line patterns from single-dish radio maps. The new approach applies linear least-squares and works on gridded maps from arbitrarily sampled data, which greatly improves computational efficiency and robustness. It also allows masking of bad data, which is useful for cases where radio frequency interference is present in the data. We evaluate the algorithms using simulations and real data obtained with the Effelsberg 100-m telescope.}


\maketitle
%
%________________________________________________________________

\section{Introduction}
In single-dish radio astronomy receiving systems often provide only one or few pixels because the necessary feed horns cannot be as densely packed as for example pixels on a CCD in optical astronomy. Therefore, to obtain sufficient mapping speed, most observers choose to observe in so-called \textit{on-the-fly} mode, also known as \textit{drift scanning}. Here, the telescope is continuously driven across the target area, forming scan lines or even just a single continuous path \citep[e.g., following a Lissajous curve; see also][]{kovacs08}. This minimizes the duty cycle and also allows very shallow observations by scanning the map as fast as necessary, constrained only by the maximal  telescope speed.

Because gain of the receiving system, the atmosphere, and contribution of the ground may vary with time, one often encounters image artifacts in the form of stripes (the scan pattern) after regularly sampling the observations on a pixel grid (gridding). One possible solution to remove such effects was proposed by \citet{haslam70}, called the basket-weaving technique, as an ``iterative process which minimized the temperature differences at crossing points''. Subsequently, this was applied to the 408-MHz survey \citep{haslam81}.

Basket-weaving refers to a strategy where the target area of interest is mapped twice with (approximately) orthogonal scanning direction. Since the observing system should receive the same contribution from the astronomical sky, the difference of both measurements should only be due to the same effects causing the image artifacts mentioned above. The aim is to find the best correction (offsets and/or gain factors) to each scan line until both maps lead to a consistent result.

For the 408-MHz survey \citet{haslam81} optimized for moderate-gain variations of the receiver using a linear function for each scan line. The optimization itself was applied to the scan lines one by one and the procedure was repeated until the residual effects were of the same order as the noise level.

There also exist other methods to deal with scanning effects that do not rely on basket-weaving. For example, \citet{sofue79} described a filtering approach (based on unsharp-masking) to suppress artifacts that usually have a different spatial scale --- relatively elongated in one dimension and smaller than the beam size in the other (at least if the maps were spatially fully-sampled). Another filtering technique was proposed by \citet{emerson88}, where  again two orthogonal maps are necessary that are combined in the  Fourier space using a weighted averaging. Both methods have in common that the spatial scales suppressed by the filter have to be known or approximated beforehand. Removing certain scales may affect parts of the astronomical signal as well. We point out that these filters can only remove additive effects, e.g. emission of the atmosphere, and not multiplicative ones, e.g. caused by gain variations or atmospheric dampening.

More recent examples of large projects that employ basket-weaving can be found in \citet{wolleben10} and \citet{peek11}. Since the data volume produced by modern receiving systems has substantially increased, it is often computationally inefficient to follow an iterative approach. This is especially the case for large-area surveys carried out with multi-feed systems, like the Effelsberg--Bonn \ion{H}{i} survey \citep[EBHIS,][]{winkel10,kerp11}, where about 320 individual scan lines (per coverage of a $5\times5\,\mathrm{deg}^2$ field) are produced. For each scan line and iteration the residual RMS had to be computed for $\sim10^4$ pixels in each of thousands of channel maps.

In this paper we present an algorithm that is based on linear least-squares for solving the basket-weaving problem directly. Furthermore, the problem matrix has to be computed only once per map, allowing one to re-use it for each channel map in the case of spectral data sets. Another clear advantage is that the algorithm operates on gridded data, such that no interpolation is needed to find the intensities on the intersection points \citep[compare][]{haslam81}. The algorithm still is only able to remove additive effects (due to the linear least-squares approach) which, in practice, rarely imposes a limitation as modern receivers are relatively stable such that gain variations are usually small. Furthermore, using calibration techniques (e.g., keeping track of the signal of a noise diode) gain effects imposed by the receiver can easily be handled. Another phenomenon with both a multiplicative (attenuation of the astronomical 
signal) and additive effect (emission of the atmosphere itself) on the astronomical signal is opacity, $\tau$, of the atmosphere. Using models \citep[e.g.,][]{pardo06} and simultaneous measurements of water vapor in the atmosphere \citep[e.g. using a water vapor radiometer,][]{wvrtechreport} one is able to determine $\tau(t)$ with sufficient accuracy to remove most of the associated gain and offsets effects. However, residual uncertainties will manifest themselves mainly as baseline offsets and can be dealt with using our approach.

Techniques very similar to basket-weaving exist. For the Planck mission \citep{tauber10}, for example, a method called \textit{de-striping} \citep[e.g.,][and references therein]{delabrouille98,efstathiou07} was used to remove the differences between the individual stripes of data obtained by observations during a slow precession of the Planck satellite. This leads to a scan geometry quite different from typical basket-weaving scan patterns. As a consequence, one ``scan line'' can have intersection points with all others, which increases the complexity of the problem even further.

The paper is organized as follows. In Section\,\ref{sec:basketweaving} we show how to solve the basket-weaving problem using a linear least-squares approach in general. As previously mentioned, one can find an implementation that operates on gridded data, which can greatly improve computational efficiency if the data were spatially oversampled (i.e., the number of intersection points is much larger than number of pixels in the map) or if spectral data with many channel maps is to be processed. This is presented in Section\,\ref{sec:griddedbasketweaving} and accompanied by simulations. Two examples using real data are discussed in Section\,\ref{sec:examples} and we conclude with a brief summary (Section\,\ref{sec:summary}).

\section{Least-squares basket-weaving}\label{sec:basketweaving}
\begin{figure}[!t]
\centering%
\includegraphics[width=0.4\textwidth,clip=]{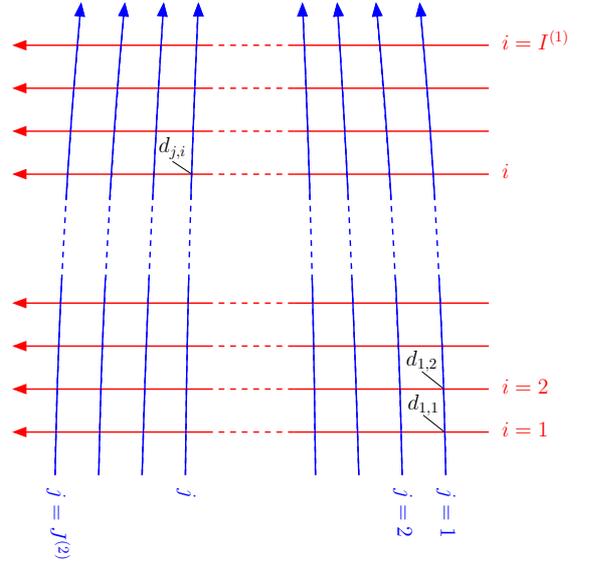}\\[0ex]
\caption{Geometry of the basket-weaving problem. Usually two (quasi) orthogonal maps are observed in so-called \textit{on-the-fly}-mode (OTF) where the telescope constantly drives across the area. Each of these \textit{scan lines} is visualized by an arrow (red: first coverage in longitudinal direction, blue: second lateral coverage). At the crossing points $d_{j,i}$ the contribution of the true brightness temperature should be the same for both coverages, while baseline (offsets) may differ. The distance between two recorded samples (dumps) on a scan line is often smaller than the separation of the scan lines. Depending on the projection and coordinate system chosen, the scan lines may not be plane-parallel. }%
\label{fig:geometry}%
\end{figure}

As discussed in the introduction, basket-weaving aims to minimize the difference, $d_{j,i}$, of the measured intensities, $y^{(1)}_{j,i}$ and $y^{(2)}_{j,i}$, at each intersection point $(j,i)$ in two quasi-orthogonal coverages. Here, $i$ and $j$ are indices over the number of scan lines, $I^{(1)}$ and $J^{(2)}$, in the first and second coverage. In the following we will omit the superscripts $(1)$ and $(2)$ when it is clear from the context which coverage is meant. In Fig.\,\ref{fig:geometry} the geometry of the basket-weaving problem is schematically shown. It is not necessary that the scan lines are plane parallel, as long as the two coverages have a sufficient number of intersection points. One could even use very complex scan patterns, for example based on Lissajous curves \citep[see also][]{kovacs08}.

Since the astronomical emission, $T_{j,i}$, should be the same in all coverages, we find
\begin{equation}
\begin{split}
d_{j,i}&\equiv y^{(1)}_{j,i}-y^{(2)}_{j,i}=T_{j,i}+p^{(1)}_{i}-(T_{j,i}+p^{(2)}_{j})\\
&=p^{(1)}_{i}-p^{(2)}_{j}~,\qquad i=1\ldots I,~j=1\ldots J. \label{eqweaver}
\end{split}
\end{equation}
The $p^{(k)}_{n}$ are the offset values for coverage $k$ and the $n$-th scan line\footnote{In Section\,\ref{subsec:polynomialoffsets} we also discuss the more general case of arbitrary (smooth) offsets which can be parametrized by polynomials.}. The latter are unknown quantities, we only know their pair-wise difference. The aim is to determine a set of $p^{(k)}_{n}$ that simultaneously fulfills all $I\times J$ equations. To enable a least-squares approach, it is useful to find a matrix notation,
\begin{equation}
\begin{split}
D_{r}&= y^{(1)}_{\lfloor r/J\rfloor,(r\bmod J)}-y^{(2)}_{\lfloor r/J\rfloor,(r\bmod J)}\\
&=\sum_{s=1\ldots I+J} (\delta_{s,(r\bmod J)}-\delta_{s,\lfloor r/J\rfloor+I})P_s\\
&\equiv\sum_{s=1\ldots I+J}  A_{r,s}P_s\label{eqweavermatrix}
\end{split}
\end{equation}
with $\vec P^T\equiv (\vec p_1^T,\vec p_2^T)=(p^{(1)}_{1},p^{(1)}_{2},\ldots,p^{(1)}_{I},p^{(2)}_{1},p^{(2)}_{1},\ldots,p^{(2)}_{J})$ and $\vec D^T\equiv (\vec d_1^T,\vec d_2^T,\ldots,\vec d_I^T)$ and $\vec d_i^T=(d_{1,i},d_{2,i},\ldots,d_{J,i})$. The notion $\lfloor x\rfloor$ is used for the floor of $x$ (the largest integer not greater than $x$). The matrix $ A$ has the size $(I\times J,I+J)$. In short,
\begin{equation}
\vec D= A \vec P\,.
\end{equation}

This linear equation can be solved using linear least-squares (e.g., via singular value decomposition, SVD) by minimizing
\begin{equation}
\left\vert A\vec P-\vec D \right\vert^2 \,.\label{eqbasketweavingleastsquares}
\end{equation}
Different noise realizations usually lead to different least-squares solutions. Furthermore, the above equation is degenerated (i.e., has no unique solution intrinsically).

\subsection{Degeneracy}\label{subsec:degeneracy}
For example, one could just add an arbitrary value $\overline p$ to all $p^{(1)}_{i}$ and $p^{(2)}_{j}$ without changing $d_{j,i}$. Therefore, it is necessary to apply additional constraints. One common way to do that is to include dampening (of the parameter vector) by minimizing the weighted sum of squared norms instead
\begin{equation}
\left\vert A\vec P-\vec D \right\vert^2 + \lambda^2\left\vert \vec P \right\vert^2,
\end{equation}
which is equivalent to
\begin{equation}
\left\vert\begin{pmatrix} A\\ \lambda\mathds{1}\end{pmatrix} \vec P-\begin{pmatrix}\vec D\\ \vec 0\end{pmatrix} \right\vert^2
\end{equation}
with the dampening value $\lambda$. This will lead to a bounded solution $\vec P$ \citep[see for example][and references therein]{Boyd04,buss09}. The choice of a reasonable value for $\lambda$ depends on the problem itself. We will discuss this briefly in Section\,\ref{subsec:choosinglambda}.

\section{Improving computational efficiency}\label{sec:griddedbasketweaving}
The method presented in Section\,\ref{sec:basketweaving} is computationally demanding, because each intersection point requires one additional row in the linear equation. Furthermore, one has to determine the intersection coordinates, $(\alpha(j,i),\delta(j,i))$, and potentially interpolate the data to obtain the values $y_{j,i}$ which usually will not match the spatial position, $(\alpha_a,\delta_a)$, of the recorded samples. 

Often the recorded data are more densely sampled than required by the Nyquist sampling theorem. For these cases we propose a modified basket-weaving algorithm that operates on gridded data. We will also show that this approach has several additional advantages (and few disadvantages). As a prerequisite, a convolution-based gridding algorithm is briefly discussed, which can be implemented to allow serial data processing (useful for large data sets) and can be easily parallelized.

The \textit{de-striping} algorithm used for the Planck mission is very similar to the least-squares technique discussed in Section\,\ref{sec:basketweaving}. It simplifies the problem of interpolating the raw data onto the intersection points by working on a special pixelization of the sky --- the HEALpix grid \citep{gorski05} --- which is well-suited for the special scan geometry.

\subsection{A convolution-based gridding method}
To grid a number of data points $y_a(\alpha_a,\delta_a)$ into a map consisting of $M\times N$ pixels $(m,n)$ with the world coordinates $(\alpha_{m,n},\delta_{m,n})$, one can apply the formula
\begin{equation}
R_{m,n}[y]=\frac{\sum_{a}y_a(\alpha_a,\delta_a)w(\alpha_{m,n}-\alpha_a,\delta_{m,n}-\delta_a)}{\sum_{a}w(\alpha_{m,n}-\alpha_a,\delta_{m,n}-\delta_a) },
\end{equation}
where $a$ is an index over the \textit{total number} of measured data points to be gridded and $w(\alpha_{m,n}-\alpha_a,\delta_{m,n}-\delta_a)$ is a weighting function (or so-called convolution kernel). For the remaining part of this paper we will use a Gaussian convolution kernel that has convenient Fourier properties but slightly degrades the spatial resolution in the resulting map\footnote{Note that the size of the kernel is loosely linked to the beam size of the instrument. The resolution in the gridded map must be good enough to ensure Nyquist sampling, but it should also be fine enough such that the gridding kernel is fully sampled. Therefore, to avoid extreme oversampling of the image, the gridding kernel should not be much smaller than the instrumental resolution. In this paper we work with a kernel having half the beam size, leading to a degradation of about 12\% of the spatial resolution in the final map.}. Now $(m,n)$ is not the crossing point of scan line $i$ (first coverage) and $j$ (second coverage), but refers to the pixel with (pixel) coordinates $(m,n)$ in the gridded map.

In the following we make use of the shorter notation $w(m,n;a)\equiv w(\alpha_{m,n}-\alpha_a,\delta_{m,n}-\delta_a)$ and $w_{m,n}\equiv \sum_{a}w(m,n;a)$.

\subsection{Re-sampling the problem matrix}\label{subsec:weavergrid}
\begin{figure}[!t]
\centering%
\includegraphics[width=0.4\textwidth,clip=]{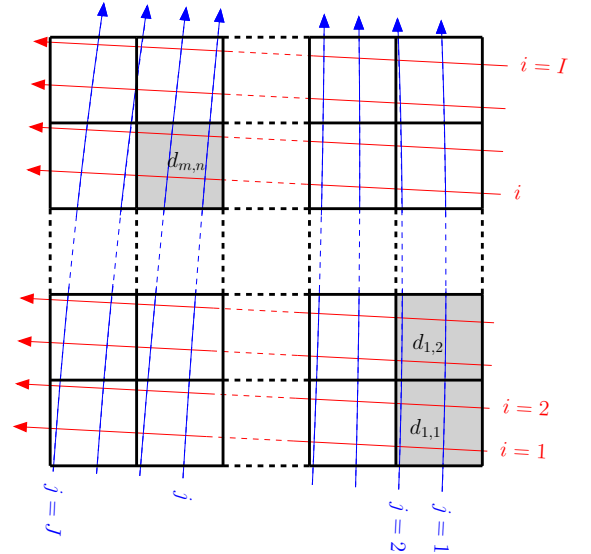}\\[0ex]
\caption{On top of the geometry sketched in Fig.\,\ref{fig:geometry} a regular pixel grid (black lines) is constructed. In this example there are fewer pixels than intersection points. The pixel grid does not need to be aligned with the underlying scan lines.}%
\label{fig:geometry2}%
\end{figure}

If the two coverages were (independently) gridded to yield the maps $R^{(1)}_{m,n}$ and $R^{(2)}_{m,n}$ we can compute a difference map (see also Fig.\,\ref{fig:geometry2} for a sketch of the geometry), similar to the approach in Eq.\,\eqref{eqweaver}
\begin{align}
d_{m,n}&\equiv  R_{m,n}[y^{(1)}]-R_{m,n}[y^{(2)}]\label{eqvecd}\\
&=R_{m,n}[T^{(1)}]+R_{m,n}[p^{(1)}]-R_{m,n}[T^{(2)}]-R_{m,n}[p^{(2)}]\\
&=\frac{1}{w^{(1)}_{m,n}}\sum_{a^{(1)}}p^{(1)}_{a^{(1)}}w(m,n;a^{(1)})-\frac{1}{w^{(2)}_{m,n}}\sum_{a^{(2)}}p^{(2)}_{a^{(2)}}w(m,n;a^{(2)}),
\end{align}
where we have used that $R_{m,n}[T^{(1)}]= R_{m,n}[T^{(2)}]$, which one can safely assume if both maps are fully sampled and the true sky brightness has not changed.

As mentioned, $a$ is the \textit{total number} of measured data points being gridded (per coverage). For later use, it obviously makes sense to also store the scan-line numbers $i$ (coverage 1) and $j$ (coverage 2) each dump $a$ belongs to, as well as the number of a dump in its scan line. The latter shall be denoted with $u$ (coverage 1) and $v$ (coverage 2). Since the number of dumps per scan line may vary, we call $U_i$ ($V_j$) the total number of dumps in scan line $i$ ($j$) in coverage 1 (2).

When we again assume that each dump in a scan line is subject to the same offset (i.e., we work with one offset per scan line), we can write
\begin{equation}
\begin{split}
d_{m,n}=&+\frac{1}{w^{(1)}_{m,n}}\sum_{i=1}^{I} p^{(1)}_i \sum_{u=1}^{U_i}w\left((m,n;a^{(1)}(i,u)\right)\\
&-\frac{1}{w^{(2)}_{m,n}}\sum_{j=1}^{J} p^{(2)}_j \sum_{v=1}^{V_j} w\left((m,n;a^{(2)}(j,v)\right)\label{eqweavergrid},
\end{split}
\end{equation}

\begin{figure*}[!t]
\centering%
\includegraphics[width=0.4\textwidth,bb=97 193 530 591,clip=]{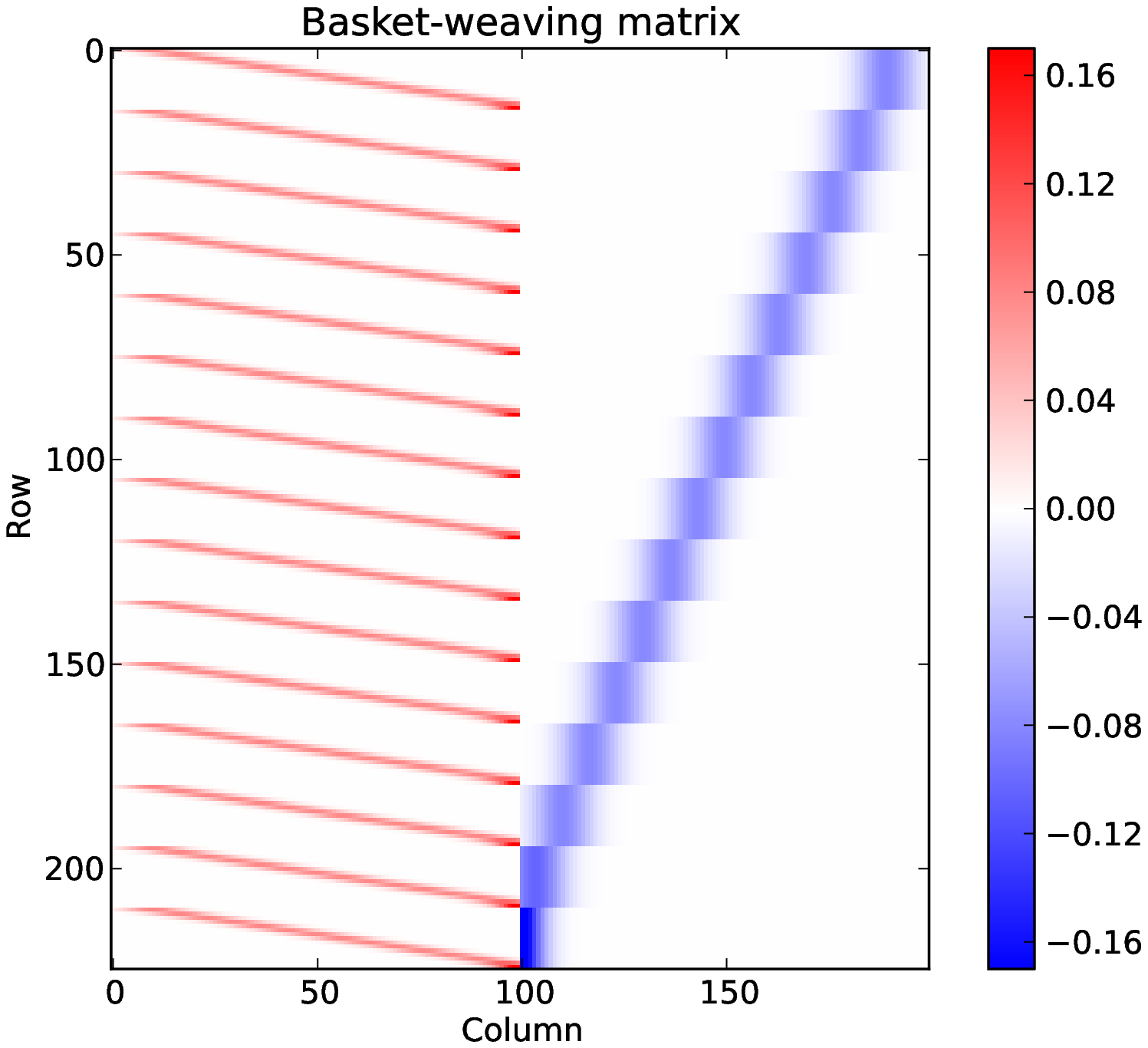}\qquad \includegraphics[width=0.4\textwidth,clip=,bb=97 193 530 591]{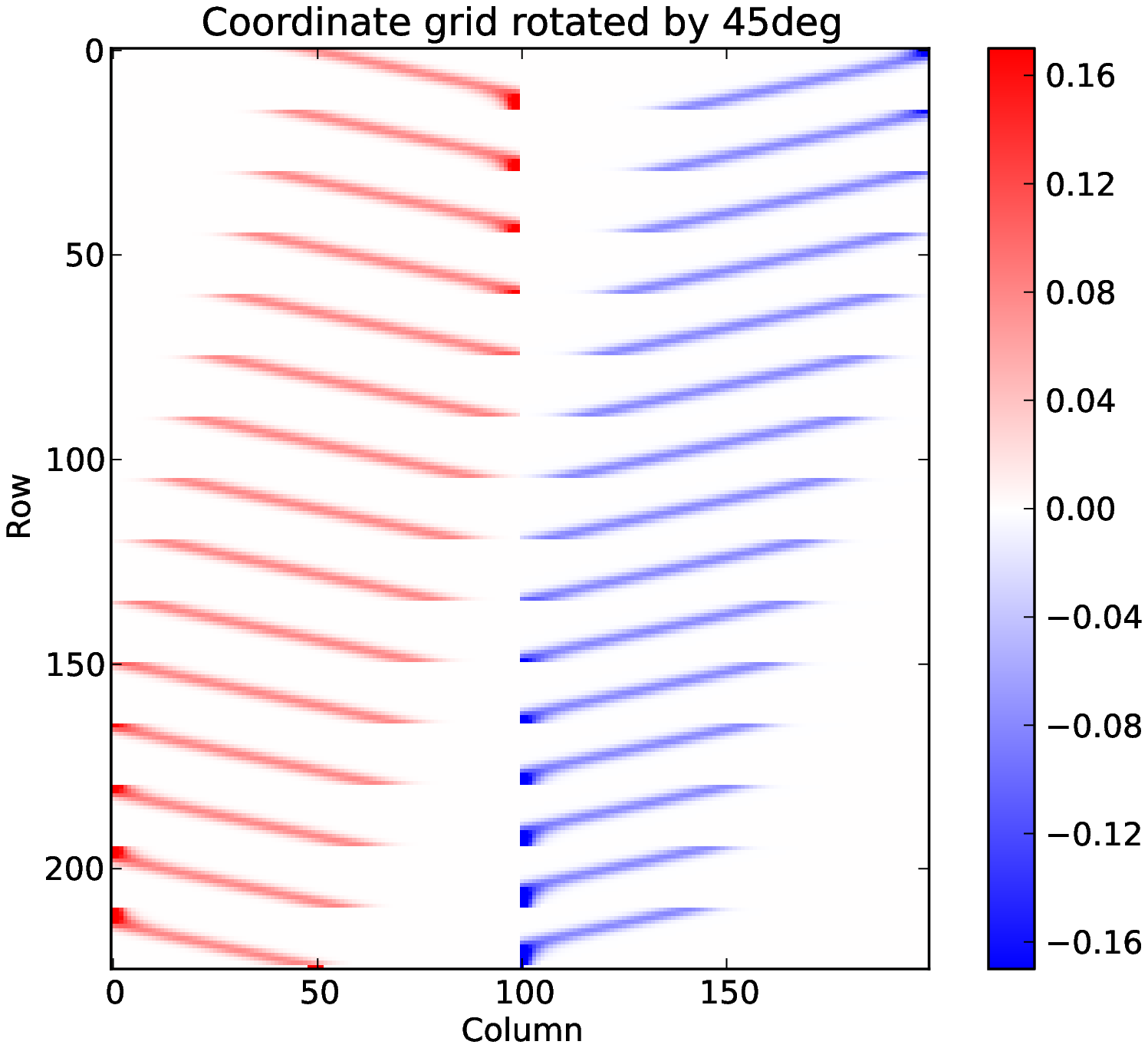}
\caption{Left panel: Basket-weaving matrix $A$ for a small test case ($2\times100$ strictly orthogonal scan lines, gridded into a map of $15\times15$ pixels). The left/right part of the matrix (columns 1 to 100/columns 101 to 200) links the offsets (100 each) associated with first/second coverage to the difference map (flattened to form the vector $\vec D$). Right panel: For the same example the coordinate grid was rotated by $45^\circ$. Evidently, now different offsets contribute to a pixel compared to the left panel.}%
\label{fig:bwmatrix}%
\end{figure*}

As before, we obtain a matrix notation by just flattening $d_{m,n}$ using the mapping $(m,n)\rightarrow r$ with $m(r)=(r\bmod M)$ and $n(r)=\lfloor r/M\rfloor$. As before, the matrix is still only dependent on the geometry of the two coverages (i.e., the sky positions at which data were taken) and the choice of a gridding kernel. 
\begin{equation}
\begin{split}
D_{r}&=\sum_{s=1\ldots I+J} \left[\Pi_1^I(s)\sum_{u=1}^{U_{i(s)}}\frac{w\left(m(r),n(r);a^{(1)}(i(s),u)\right)}{w^{(1)}_{m(r),n(r)}}\right.\\
&\left.-\Pi_{I+1}^{I+J}(s)\sum_{v=1}^{V_{j(s)}}\frac{w\left(m(r),n(r);a^{(2)}(j(s),v)\right)}{w^{(2)}_{m(r),n(r)}}\right]P_s\\
&\equiv\sum_{s=1\ldots I+J}  A_{r,s}P_s\label{eqweavergridmatrixequation}
\end{split}
\end{equation}
with
\begin{equation}
\Pi_a^b(s)\equiv\begin{cases}1\quad:~a\leq s \leq b\\ 0\quad:~\mathrm{else}\end{cases}
\end{equation}
and $i(s)$ and $j(s)$ being the mapping functions to obtain from the sequence ${s}$ the appropriate scan line number for the two coverages, i.e., $i(s)=s$ for $1\leq s\leq I$ and $j(s)=s-I$ for $I+1\leq s\leq I+J$. Again, $P_s$ is the vector containing all offsets as defined in Section\,\ref{sec:basketweaving}.

The two necessary quantities, $\vec D$, and the matrix, $ A$ can be easily obtained using the following recipe
\begin{enumerate}
\item Grid all data points of coverage 1 and 2 into two separate maps, $R^{(1,2)}_{m,n}$ and keep the weight maps $w^{(1,2)}_{m,n}$.
\item Subtract both maps according to Eq.\,(\ref{eqvecd}) and flatten the result to obtain $\vec D$.
\item During the gridding process, compute additional $I+J$ maps: for each of the $I$ ($J$) scan lines in coverage 1 (2), set all data points to 1 and grid separately each scan line into one individual map. Divide the first $I$ maps by $w^{(1)}_{m,n}$ and the following $J$ maps by $w^{(2)}_{m,n}$. 
\item Flatten each of the resulting $I+J$ weight maps; the resulting vectors form the columns of the matrix $A$.
\end{enumerate}

While equation Eq.\,(\ref{eqweavergrid}) looks more complicated than Eq.\,(\ref{eqweaver}), the problem matrix is just a ``smoothed'' version of the former, having the same basic structure. Likewise, the calculated offset values $P_s$ are just a ``smooth'' estimate of the true offset values, having the same effect on the \textit{gridded} data. The ``re-sampled'' matrix $A$ has $M\times N$ rows and $I+J$ columns, i.e., it can be of much smaller size if the the original maps were over-sampled. 

In Fig.\,\ref{fig:bwmatrix} we have visualized the basket-weaving matrix for a small example. Each row in the matrix corresponds to one pixel in the difference map, i.e. it links various scan line offsets (having different weights) to it. The left part of $A$ refers to the first coverage, while the right part refers to the second (here values in $A$ are negative to achieve subtraction of both coverages). The example in the left panel of Fig.\,\ref{fig:bwmatrix} is for strictly orthogonal scan lines. To demonstrate a more complicated case, we simply rotated the coordinate grid by $45^\circ$ (but not the scan lines), leading to the matrix in the right panel of Fig.\,\ref{fig:bwmatrix}.

As before, additional constraints are necessary (e.g., the dampening strategy), as the problem is still degenerated.

\subsection{Calculating the correction map}
Using the gridding-based approach facilitates calculating a correction map without the need to re-run the gridding process (with the computed offsets applied). This is possible because the offsets are additive, and the matrix $A$ already contains all necessary weight factors. We find that
\begin{equation}
\begin{split}
C_{r}&=\frac{\sum_{s=1}^{I} w^{(1)}_{m(r),n(r)}A_{r,s} P_s + \sum_{s=1}^{J} w^{(2)}_{m(r),n(r)}\vert A_{r,s}\vert P_s}{w^{(1)}_{m(r),n(r)}+w^{(2)}_{m(r),n(r)}}\\
&=\frac{\sum_{s=1}^{I+J} \left[\Pi_1^I(s)w^{(1)}_{m(r),n(r)}+\Pi_{I+1}^{I+J}(s)w^{(2)}_{m(r),n(r)}\right]\vert A_{r,s}\vert P_s}{w^{(1)}_{m(r),n(r)}+w^{(2)}_{m(r),n(r)}}\label{eqcorrectionmap}
\end{split}
\end{equation}
is the overall correction map (after re-shaping to two dimensions)\footnote{Taking the absolute value of the elements is necessary since the matrix $A$ contains negative values in the columns $I+1$ to $I+J$.}. If one similarly applies $\vert A\vert$ to subsets of $\vec P$, one can also obtain correction maps for the individual coverages.

\begin{figure*}[!t]
\centering%
\includegraphics[width=0.98\textwidth,bb=-55 291 657 517,clip=]{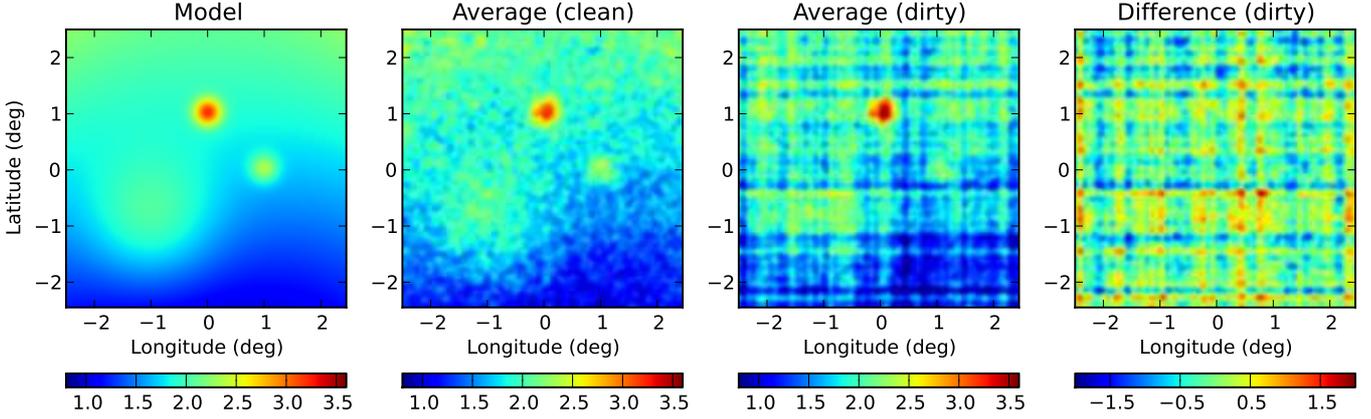}\\[0ex]
\caption{Simulated maps. The left panel contains the input sky model. From this, two coverages were drawn and noise was added as well as a random offset for each scan line. The middle panels show the average of both maps (for the clean and dirty case, the former with and the latter without offsets applied), the right panel the difference $d_{m,n}$ according to Eq.\,\eqref{eqvecd}. Note that different intensity scales were used.}%
\label{fig:simulationinput}%
\end{figure*}

\begin{figure*}[!t]
\centering%
\includegraphics[width=0.98\textwidth,bb=-55 291 657 517,clip=]{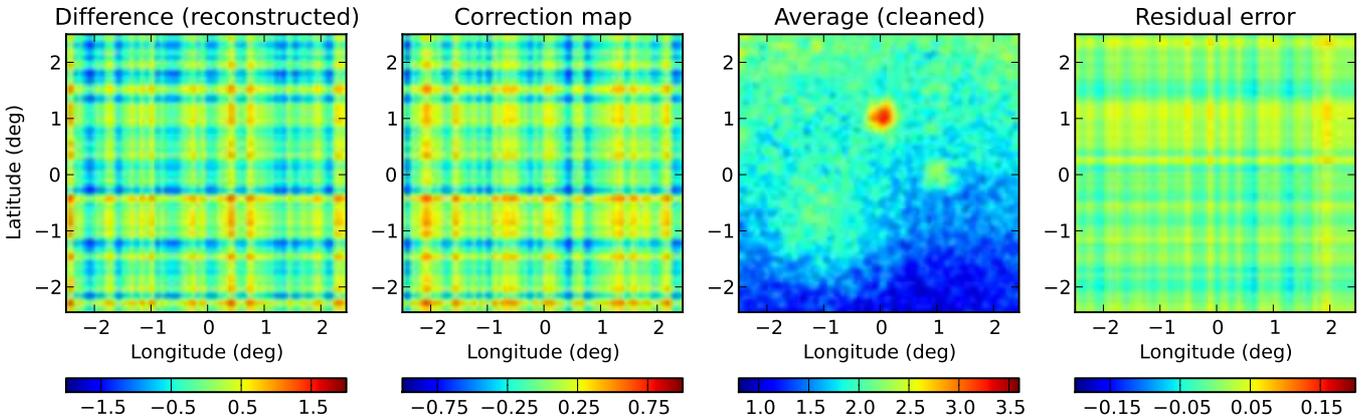}\\[0ex]
\caption{Solving Eq.\,\eqref{eqweavergridmatrixequation} for the difference map $d_{m,n}$ (Fig.\,\ref{fig:simulationinput} right panel), one obtains a solution for the offset values $\vec P$. For comparison we show in the left panel the reconstructed difference map (which is basically a noise-free version of the input difference map). Via Eq.\,\eqref{eqcorrectionmap} one can easily compute a correction map (second panel) that can be subtracted from the original ``dirty'' map (see Fig.\,\ref{fig:simulationinput} third panel). The result is a cleaned data set (third panel), from which image artifacts are visually removed. The right panel shows the difference of the clean and cleaned map. Residual stripes appear, but they are suppressed by about an order of magnitude compared to the correction map (note the different intensity scale).}%
\label{fig:simulationoutput}%
\end{figure*}

This makes the proposed method extremely useful when a large set of observations with identical spatial positions of the data samples have to be processed, as for example in the case of spectroscopic observations or state-of-the-art continuum measurements (with many channels). For these it is likely that the offsets $\vec P$ will be a function of frequency, such that basket-weaving has to be applied to many spectral channel maps. Then the matrix $A$ is the same for all channel maps, only $\vec D$ needs to be computed for each channel, which is usually done in the gridding process anyway.

\subsection{Simulations}\label{subsec:simulations}
We performed simulations to test the proposed basket-weaving method that also allowed us to analyze the quality of the solution and the performance.  The simulated map parameters resemble those of the Effelsberg--Bonn HI Survey \citep[EBHIS,][]{winkel10,kerp11} since this is one of the potential applications of least-squares basket-weaving. For EBHIS the angular resolution is 10\arcmin, which means that for a typical field size of $5\degr\times5\degr$ one needs about $100\times100$ pixels of size $3\arcmin$ to achieve a fully sampled map after gridding. The number of observed scan lines is higher with about 320 scan lines per coverage, which means that the number of intersection points is an order of magnitude larger than the number of pixels in the final map. Each scan line contains about 160 data points (dumps).

For our simulations we used a simple sky model containing several Gaussians on top of a (2-D) polynomial background; see Fig.\,\ref{fig:simulationinput} (left panel). From this sky model we sampled two maps, one with longitudinal and one with lateral scan direction. To both of these coverages we added Gaussian noise and an offset to each scan-line, the amplitudes of which were  sampled from a Gaussian distribution (with similar standard deviation as the noise distribution). Figure\,\ref{fig:simulationinput} shows the average (middle panels) and difference (right panel) maps, clearly containing a strong pattern caused by the offsets.

After constructing and solving the matrix equation Eq.\,\eqref{eqweavergridmatrixequation} based on the difference map in Fig.\,\ref{fig:simulationinput} (right panel), one obtains a least-squares solution for the offsets $\vec P$. Multiplying the matrix $A$ with $\vec P$ leads to the reconstructed difference map (Fig.\,\ref{fig:simulationoutput}, left panel), which is basically a noise-free version of the former. Likewise, with Eq.\,\eqref{eqcorrectionmap} we can compute a correction map (Fig.\,\ref{fig:simulationoutput}, second panel) that leads to a cleaned average of both coverages (Fig.\,\ref{fig:simulationoutput}, third panel) when subtracting from the dirty average map (Fig.\,\ref{fig:simulationinput}, third panel). In this cleaned map, visually no image artifacts (stripes) appear. However, the difference between the clean and cleaned data shows some residual errors (Fig.\,\ref{fig:simulationinput}, right panel), which are about an order of magnitude smaller than the original image artifacts/stripes.

\subsection{Polynomial offsets}\label{subsec:polynomialoffsets}
Up to now we only accounted for the relatively simple case of constant offsets for each scan line. Of course, in reality it might well be that there is some underlying trend in the data of a scan line, e.g. due to weather effects or variable radiation from the ground. Then one possibility that still allows a linear least-squares approach is to parametrize the drifts using polynomials. 

Usually, individual data samples in a scan line will have the same integration time, such that one could construct the polynomials as a function of dump number per scan line. To obtain a similar order of magnitude for all polynomial coefficients (up to the maximum order of $o_p$), it is possible to re-scale the \textit{integer} dump number $1\leq u\leq U_i$ (and $1\leq v\leq V_j$) to the interval $[0,1]$ by dividing the dump number by the total number of dumps in a scan line. Eq.\,\eqref{eqweaver} then reads as 
\begin{equation}
d_{i,j}=\sum_{o=0}^{o_p^{(1)}}p^{(1)}_{o,i}\left(\frac{u_i}{U_i}\right)^o-\sum_{o=0}^{o_p^{(2)}}p^{(2)}_{o,j}\left(\frac{v_j}{V_j}\right)^o,\label{eqweaverporder}
\end{equation}

However, with this approach one has to parameterize and interpolate the position of each data point on the scan lines to obtain $u_i$ and $v_j$, \textit{which usually are not integer numbers}. This adds complexity to the problem. With the gridding-based method, this will naturally be solved, since the gridding is equivalent to interpolation.

One can easily convert Eq.\,\eqref{eqweaverporder} into a matrix equation similar to Eq.\,\eqref{eqweavermatrix}. The matrix $A$ will then not only contain entries $+1$ and $-1$ but also the values of $(u_i/U_i)^o$ and $(v_j/V_j)^o$, respectively. It has $I\times J$ rows and $(o_p^{(1)}+1)\times I+(o_p^{(2)}+1)\times J$ columns.

Likewise, the gridding-based approach can be adapted to allow for higher polynomial orders. Here, we omit the (lengthy) equations, but just extend the recipe given in Section\,\ref{subsec:weavergrid}. Only step (3) needs to be slightly changed

\begin{enumerate}
\setcounter{enumi}{2}
\item In the gridding process, compute additional $(o_p^{(1)}+1)\times I+(o_p^{(2)}+1)\times J$ maps: for each of the $I$ ($J$) scan lines in coverage 1 (2), set the data points to $(u/U_i)^o$ and $(v/V_j)^o$, respectively\footnote{Here $u$ and $v$ are again the \textit{integer} dump numbers and not interpolated quantities.}, and grid these data separately for each scan line and polynomial coefficient into one individual map. Divide the $(o_p^{(1)}+1)\times I$ maps associated with the first coverage by $(o_p^{(1)}+1)w^{(1)}_{m,n}$ and the other $(o_p^{(2)}+1)\times J$ maps by $(o_p^{(2)}+1)w^{(2)}_{m,n}$. 
\end{enumerate}

Again, the resulting matrix will have the same number of columns as for the non-gridding-based approach, i.e., $(o_p^{(1)}+1)\times I+(o_p^{(2)}+1)\times J$, but a different number of rows, $M\times N$.

Note that also the reconstruction formula must be slightly changed to account for the increased number of parameters that contribute to each pixel in the map. This is done by simply changing the denominator of Eq.\,\eqref{eqcorrectionmap} to $(o_p^{(1)}+1)w^{(1)}_{m,n}+(o_p^{(2)}+1)w^{(2)}_{m,n}$.

Of course, one cannot solve the system for an arbitrarily high polynomial degree. Artificially increasing $M$ and $N$ (the spatial resolution of the gridded map) will not help either, because adjacent pixels will be correlated and not add information. In total $M\times N$ should be much larger than $I+J\gtrsim M+N$, which just means that a sufficient size of the map is necessary\footnote{The equality is valid if the scan lines were perfectly Nyquist sampled. In all other (oversampled) cases, the grid-based approach will reduce the size of the matrix and increase computational efficiency.}.

One could also choose a completely different set of basis functions to parametrize the offsets, not only simple polynomials. Furthermore, it is possible to choose different free parameter(s), e.g., elevation instead of dump number. Both are possible because one just needs to grid the values of the functions for each coefficient into the matrix --- the structure of the basket-weaving equation is still the same.

\begin{figure*}[!tp]
\centering%
\includegraphics[width=0.98\textwidth,bb=-170 -201 778 964,clip=]{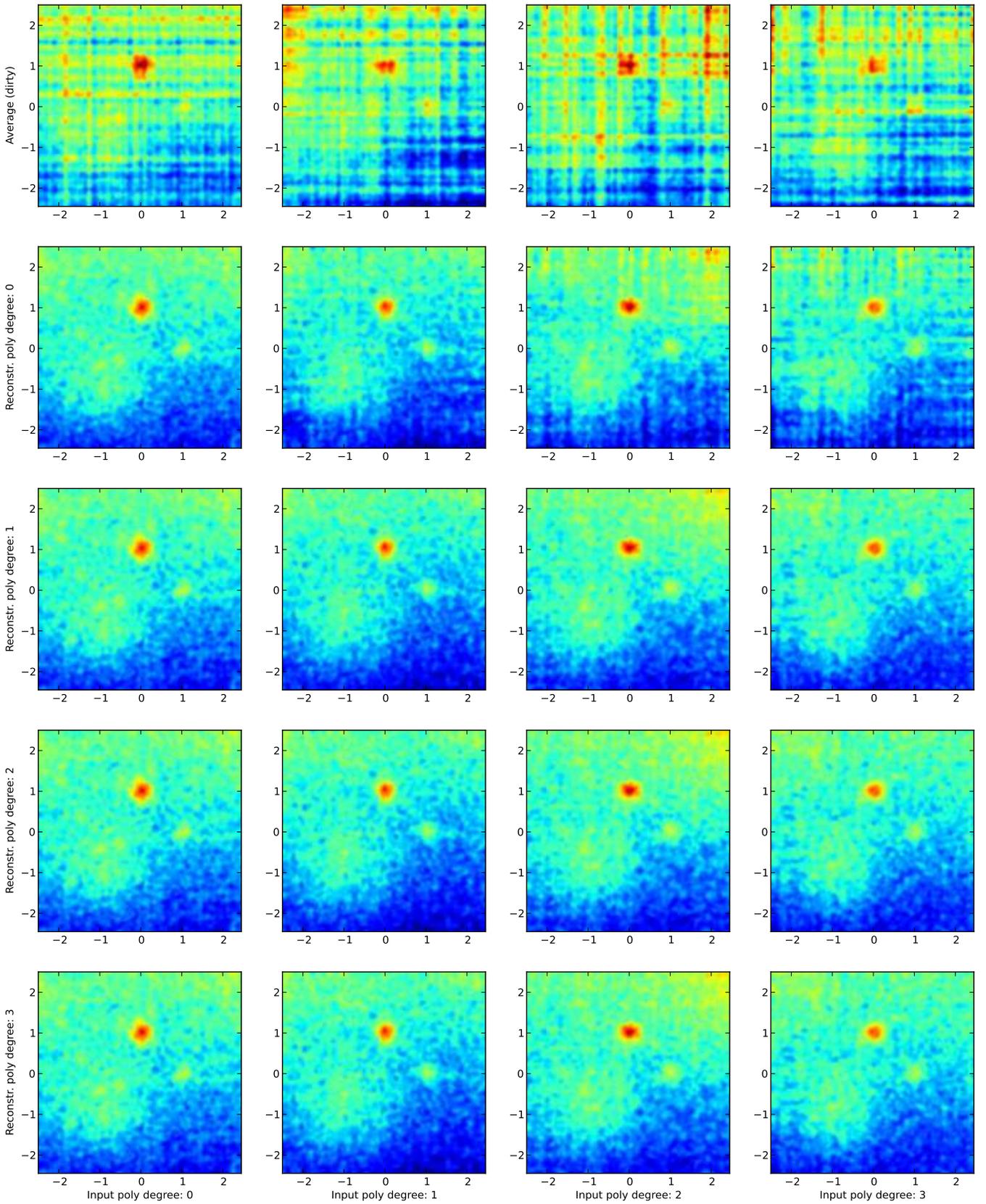}
\caption{Result of the least-squares basket-weaving correction for different polynomial degrees. The four columns belong to the generated polynomial offsets (from constant offsets in the left column to a third-order polynomial in the right column). The top row shows the dirty maps for these. The four bottom rows refer to different reconstruction degrees, i.e., the diagonal of the lower $4\times4$ matrix contains the cases where the reconstruction polynomial degree matches the simulated degree.}%
\label{fig:simulationpinpout}%
\end{figure*}

In Fig.\,\ref{fig:simulationpinpout} we show results for higher-order polynomial offsets. In reality it is not necessarily clear what the best reconstruction polynomial degree would be. Therefore, not only the cases where the simulated polynomial offsets match the polynomial degree used for the basket-weaving matrix are shown. For example, the first column shows the results for using various reconstruction degrees, all applied to data containing constant offsets for each scan line. Due to the increasing number of parameters --- using a third-order reconstruction polynomial requires already $\sim2500$ parameters, which is comparable to the total number of pixels ($10^4$) --- the solution becomes increasingly worse. The same is true if higher-degree polynomials are used to generate the data (from left to right in each row), but the reconstruction polynomial order is not sufficient to describe these. But also along the diagonal, where both polynomial degrees match, the results become worse for 
increasing degrees. Again, this is a question of number of free parameters, which are less constrained to the lower right (the noise level was kept constant for all cases).

Nevertheless, in all cases the results are still better than without basket-weaving (the top row of Fig.\,\ref{fig:simulationpinpout} contains the dirty images). 

\subsection{Choice of the dampening parameter $\lambda$}\label{subsec:choosinglambda}
\begin{figure}[!t]
\centering%
\includegraphics[width=0.45\textwidth,bb=17 4 525 392,clip=]{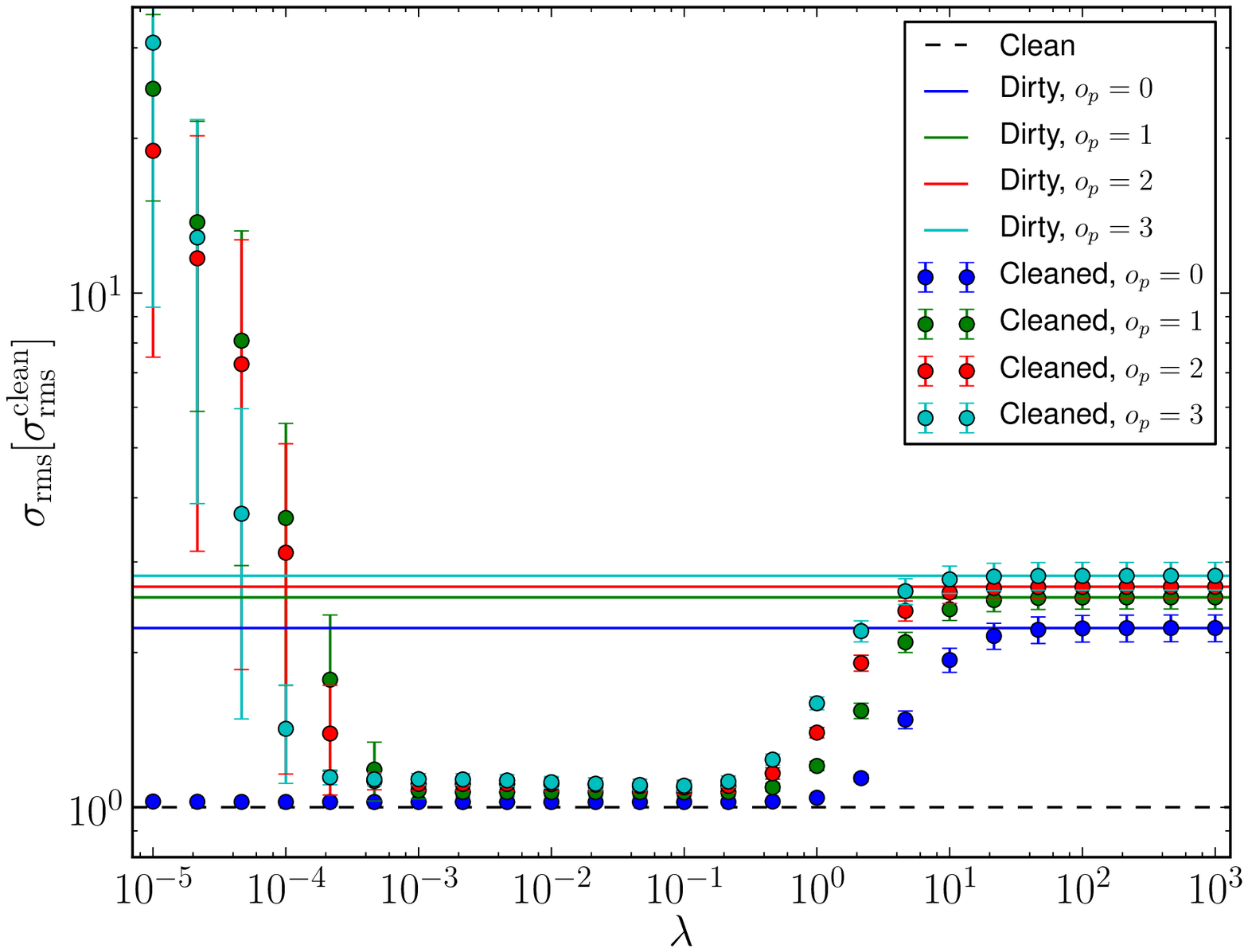}\\[0ex]
\caption{Influence of the dampening parameter $\lambda$ on the quality of the cleaned maps.}%
\label{fig:lambdarms}%
\end{figure}

In Section\,\ref{subsec:degeneracy} we discussed how the (degenerated) basket-weaving problem can be solved by applying a dampening scheme that constrains the solution. However, it is not directly evident how large $\lambda$ should be. For our example, using first-order polynomials, we processed the data with a wide range of different $\lambda$ values. To improve statistics, this was repeated 30 times, each time using different noise realization and offset coefficients. The input model was subtracted from each reconstructed map to calculate the residual RMS, which we plot in Fig.\,\ref{fig:lambdarms} (normalized to the RMS in the clean data set residuals). For reference, the plot also contains a horizontal line (green) for the value of One and the normalized residual RMS for the dirty maps. The former defines the noise level for perfect reconstruction, the latter resembles the case where no improvement was achieved compared to the dirty maps.

Fig.\,\ref{fig:lambdarms} reveals that a wide range of $\lambda$ values (three orders of magnitude) leads to a consistent solution. For large $\lambda$, which means that the constraint on the offsets is tighter, the residual RMS converges to the value of the dirty images. Here, the solution is in fact bound to zero. Toward small $\lambda$ the residual RMS constantly increases, a sign for divergence of the solution because the degeneracy of the problem is not sufficiently suppressed by the dampening anymore.

In the regime where the RMS is minimal we observe an increase of the RMS of about 2.5\% (7\%, 10\%, 12\%) for $o_p=0$ ($o_p=1,2,3$) compared to the clean maps. Considering the fact that the offset coefficients were drawn from a Gaussian noise distribution with the same standard deviation as the noise in the Raw data this is acceptable\footnote{For reference: $\sigma_\mathrm{rms}$ in the raw data was set to One resulting in $\sigma_\mathrm{rms}\approx0.1$ in the gridded data.}.

\section{Examples}\label{sec:examples}

In this section we will present two examples based on measurements with the Effelsberg 100-m telescope. 

\subsection{Continuum maps (6-cm)}
\begin{figure*}[!t]
\centering%
\includegraphics[width=0.48\textwidth,bb=15 189 550 573,clip=]{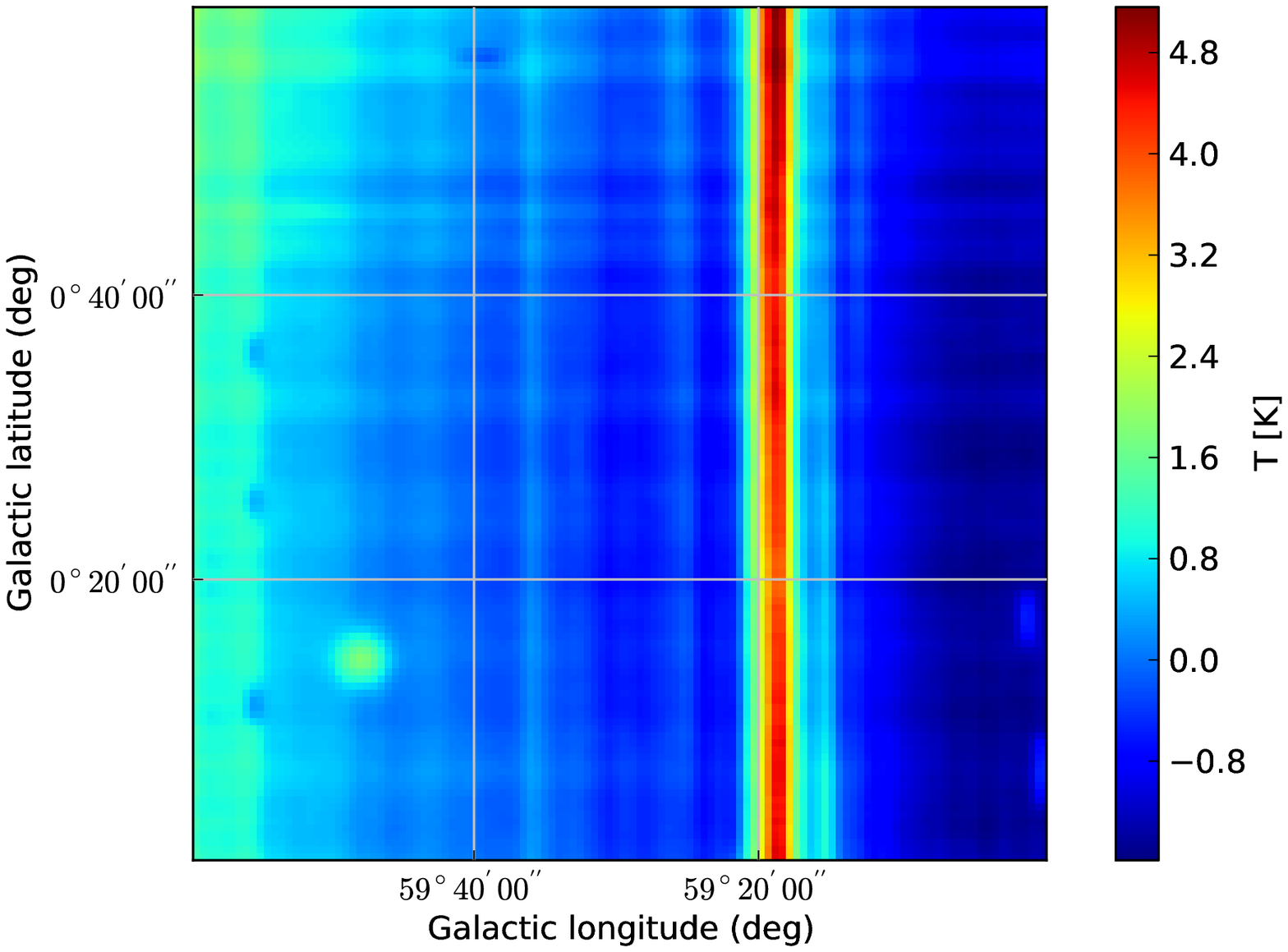}~\includegraphics[width=0.48\textwidth,bb=15 189 550 573,clip=]{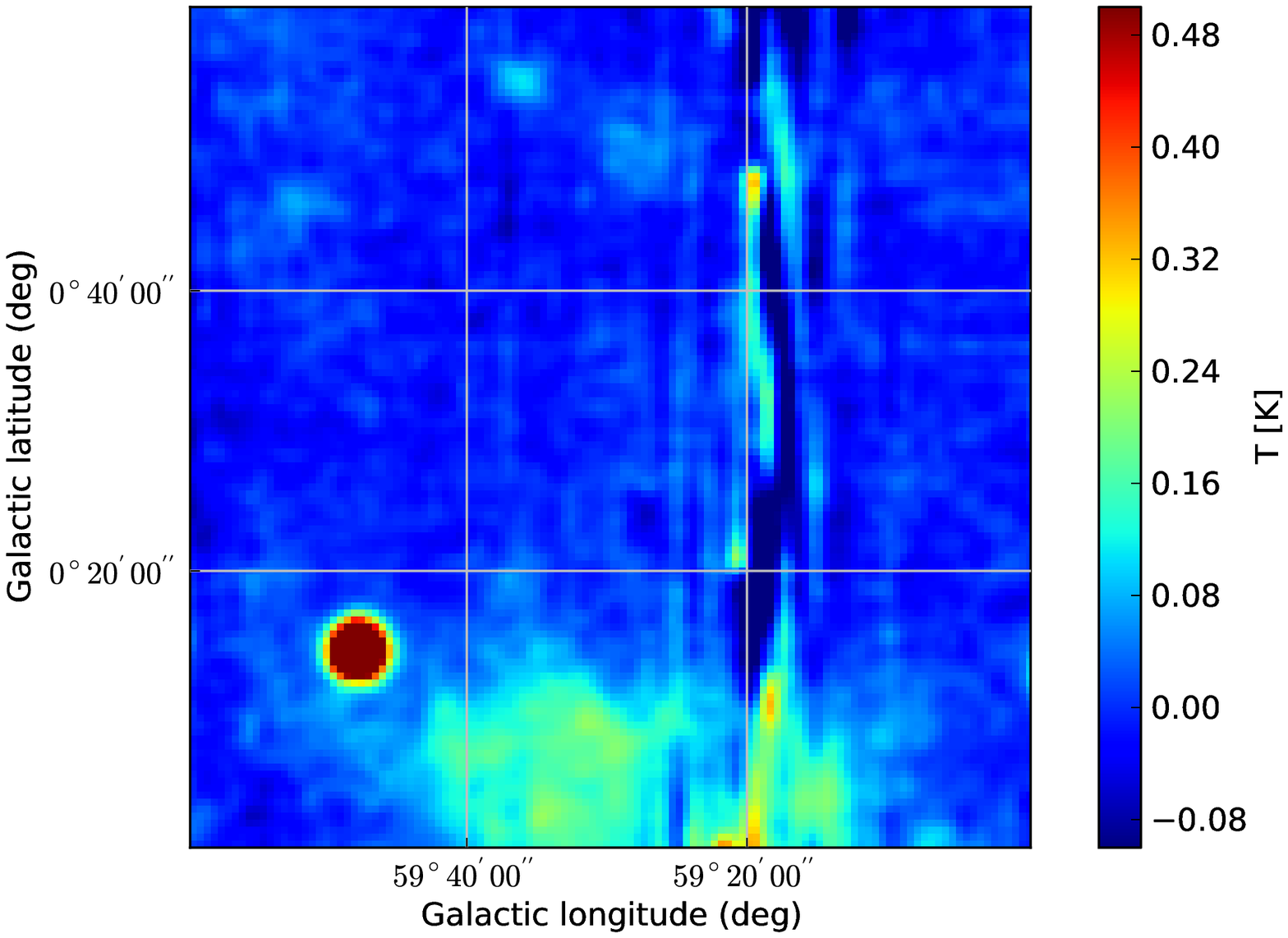}
\caption{Left panel: Map showing 6-cm continuum emission (after full calibration) of the northern part of the dark cloud G\,59.5$-$0.2. In the left panel both coverages where simply gridded together. The right panel displays the result of our basket-weaving algorithm applied. Apparently a strong residual artifact remains, which we attribute to the RFI signal at $l\approx59.3^\circ$ visible in the left panel.}%
\label{fig:6cm_original_and_after_bw}%
\end{figure*}

The first examples is taken from to a pilot study to prove the feasibility of a large spectral line and continuum survey of the Galactic plane in the 4--8\,GHz frequency range to be undertaken with the Jansky VLA. Supplementary Effelsberg 100-m data are used for short-spacings, especially for the continuum. Since currently no receiver at the 100-m covers the 4--8\,GHz range simultaneously for the tests a 6-cm receiver was used providing 500\,MHz of instantaneous bandwidth. Working in spectral-line mode, the new XFFTS \citep{klein12} backend with its high number of spectral channels (32k) provides not only the possibility to map narrow absorption and emission line features but also to extract the continuum.

For our test observations we targeted the dark cloud G\,59.5$-$0.2. Two $1\times1\,\mathrm{deg}^2$ fields were independently observed. Zig--zag on-the-fly mapping was combined with the so-called position--switching technique where, after few scan lines, the telescope was pointed to a reference position. The latter is used to remove bandpass effects. To account for frequency-dependent system temperature, $T_\mathrm{sys}$, and calibration normal, $T_\mathrm{cal}$ (provided by a noise diode), we used the method proposed in \citet{winkel12}. Accordingly, every two hours a bright continuum source of well-known flux was observed as calibration source to determine $T_\mathrm{cal}(\nu)$. Furthermore, corrections for atmospheric opacity and elevation-dependent telescope efficiency (\textit{gain curve}) were applied.

Unfortunately, during several of the scans strong broad-band radio frequency interference (RFI) was entering the receiving system, the effect of which can be seen in Fig.\,\ref{fig:6cm_original_and_after_bw}. Here, the calibrated data of the northern  field was gridded (both coverages). The strong vertical artifact at $l\approx59.3^\circ$ is caused by this RFI. But even in the remaining data there is not much to be seen, except for a bright continuum source in the lower left part of the figure. 

Applying the basket-weaving technique (using second-order polynomials) as proposed in Section\,\ref{subsec:polynomialoffsets} leads to the map shown in the right panel of Fig.\,\ref{fig:6cm_original_and_after_bw}. While the scan pattern is largely removed and some diffuse flux in the bottom part becomes visible, there is still residual RFI visible.

\begin{figure}[!t]
\centering%
\includegraphics[width=0.45\textwidth,bb=55 -59 555 835,clip=]{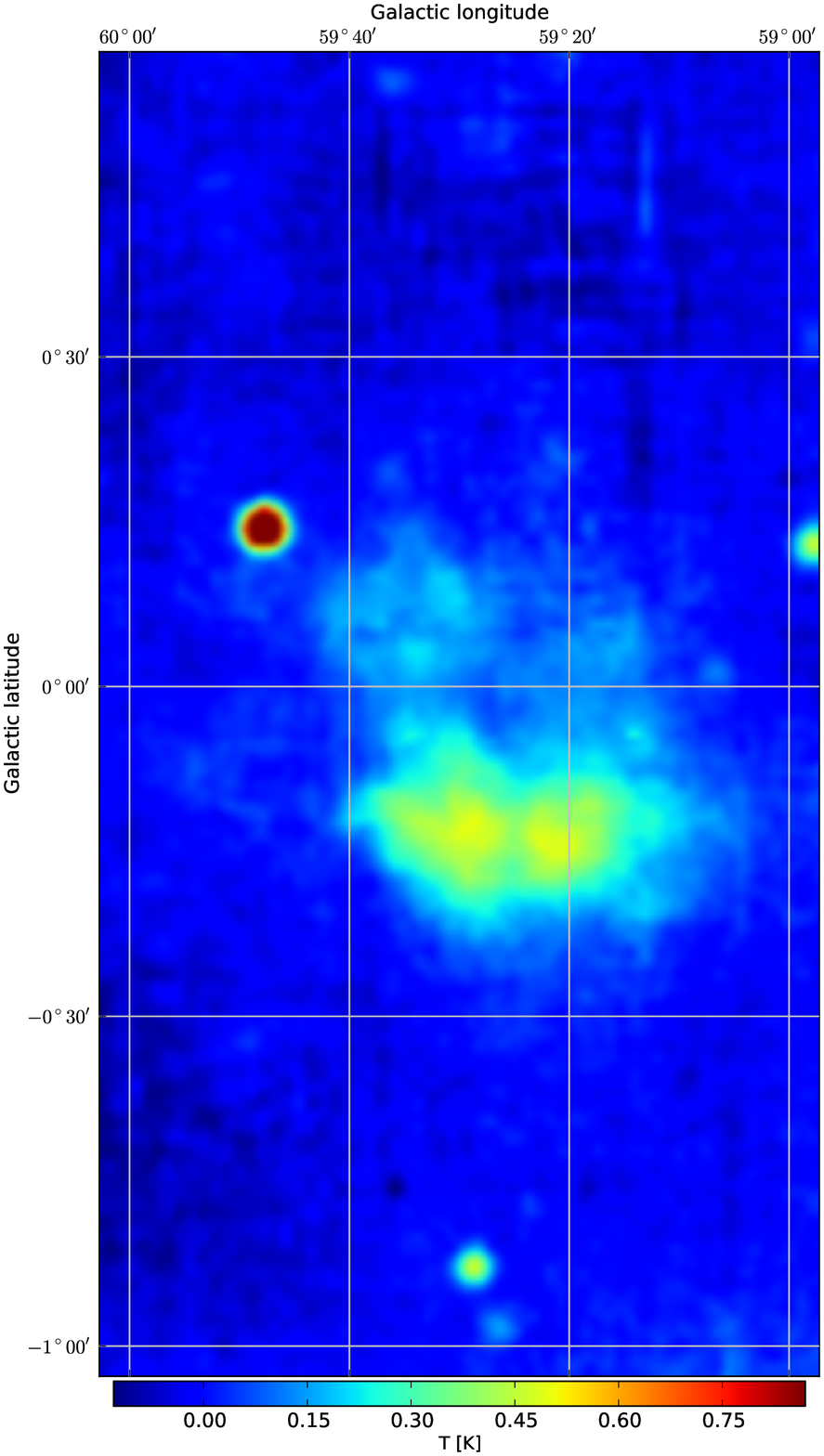}
\caption{Final 6-cm (4.8-GHz) continuum map of dark cloud G59.5$-$0.2. It was simply concatenated from the northern and southern fields, which were independently calibrated and corrected using the basket-weaving technique.}%
\label{fig:6cm_NEandSE}%
\end{figure}

To further improve the data quality, we finally masked several scan lines containing the strong RFI signals. This is simply done by removing associated rows from the solution matrix, $A$, and difference vector, $D$. (Of course, to calculate the correction map, the full matrix, $A$, needs to be used.) Furthermore, for the final gridding bad data were omitted as well. The final image composed of the northern and southern field is presented in Fig.\,\ref{fig:6cm_NEandSE}. The diffuse continuum of the dark cloud smoothly extends from the northern to the southern part, although no cross-calibration was applied between the two fields. The masking suppresses the residual RFI in the northern part effectively, but neglecting (a substantial amount of) data comes at a price: the region that was masked has a somewhat stronger remaining scan pattern simply because the solution is much less constrained there.

\subsection{Continuum maps (21-cm, EBHIS)}
For the second example we used data of the Effelsberg--Bonn \ion{H}{i} Survey \citep{winkel10,kerp11}, which is a spectroscopic survey of the neutral atomic hydrogen mapping the full northern hemisphere. It uses a seven-feed receiver with 100\,MHz instantaneous bandwidth. A total of 14 FFT spectrometers (16k spectral channels each) allow one to observe the band with sufficient spectral resolution to not only detect extra-galactic sources out to a redshift of $z\sim0.07$, but also map the \ion{H}{i} in the Milky Way. The first coverage is nearly completed.

Here we show that the data can in principle be used to recover the continuum emission as well. For a few fields we supplemented the EBHIS data with measurements in the orthogonal scanning direction. Data were reduced with the standard EBHIS pipeline \citep{winkel10} except for baseline fitting (usually employed to remove continuum from the spectral data). 

We extracted the continuum level at 21-cm by calculating the median over the whole bandwidth. This was done to be robust against radio frequency interference (RFI) that is present in roughly 10\% of the channels and can greatly exceed the average system temperature.

The system temperature at 21\,cm is strongly dependent on elevation (mostly due to atmospheric opacity but with a significant amount of radiation from the ground). This contribution has to be removed first, because the basket-weaving alone will lead to reconstructed large-scale flux including any baseline contribution (atmosphere and ground radiation) \textit{present in both maps}. Since the maps are observed in the equatorial coordinate system (scanning direction in right ascension and declination), there is usually a large-scale gradient present in both coverages as the target area moves across the (horizontal) sky.

\begin{figure}[!t]
\centering%
\includegraphics[width=0.48\textwidth,bb=40 129 539 630,clip=]{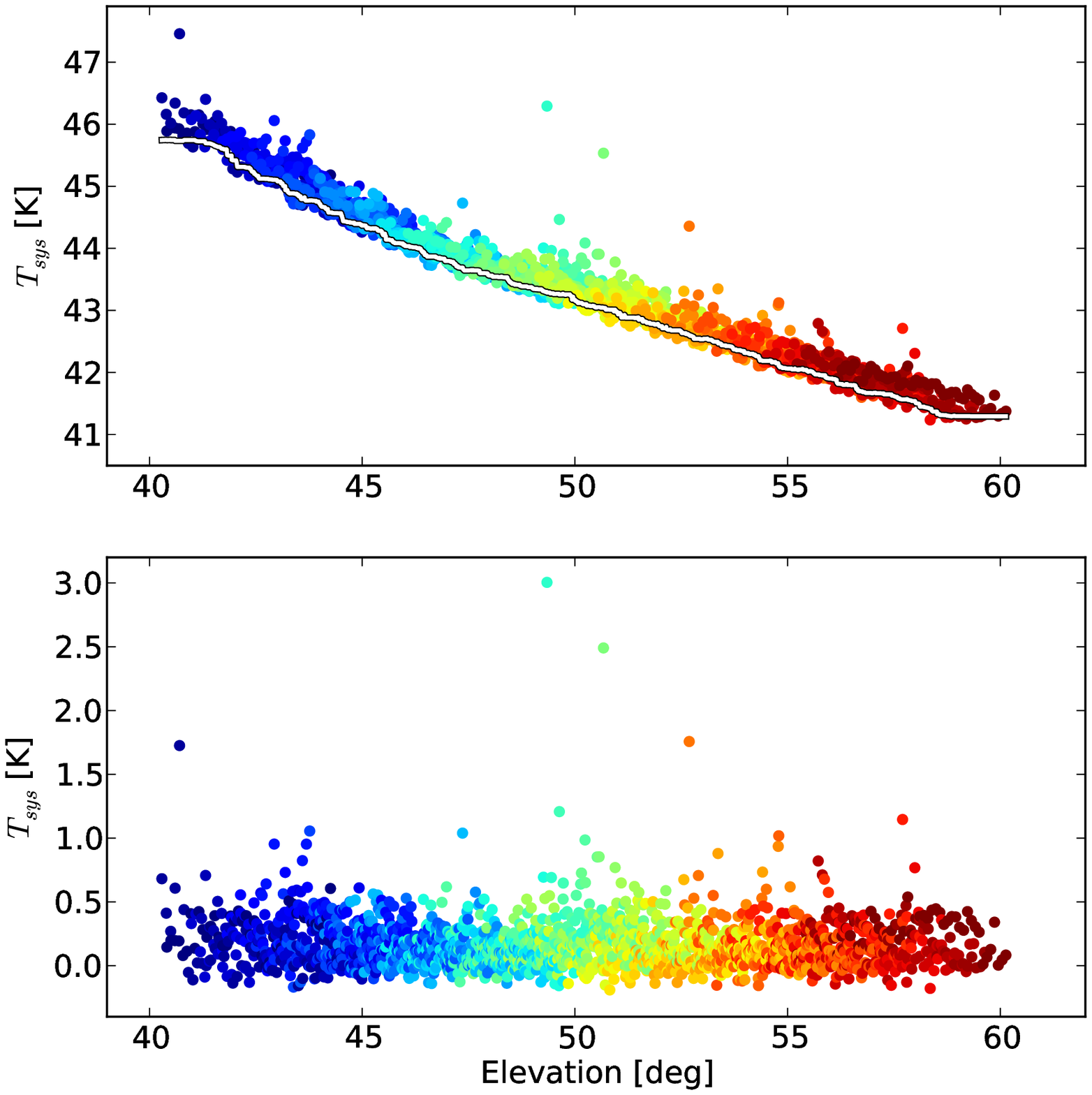}
\caption{Upper panel: the system temperature of a single feed as a function of elevation. To illustrate that multiple scan lines contribute at a given elevation, each scan line was color-coded. The line shows the result of the percentile filter (see text). Lower panel: the residual brightness temperatures after the subtraction of the percentile filter. No dependence on elevation is evident.}%
\label{fig:ebhisazfit}%
\end{figure}

While the atmospheric effect can be modeled relatively easily, the ground contribution is a more complex function of the telescope aperture and the shape and temperature of the horizon, we decided not to fit a certain model to the data but instead use a simpler approach by using a percentile filter. For each of the seven feeds, the brightness temperatures were sorted with ascending elevation (Fig. \ref{fig:ebhisazfit}, upper panel) and a running percentile filter, which calculates the 15\% percentile from 50 values, was applied. This filtered version, which is a lower envelope to the measured system temperatures, was then used to remove the shape and amplitude of the ground level (Fig. \ref{fig:ebhisazfit}, upper panel, line). Although the residual brightness temperatures (Fig. \ref{fig:ebhisazfit}, lower panel) have also lost information about the absolute continuum level, they should still contain information on the shape of the overall sky distribution, since the subtraction was made independently of sky position.

\begin{figure*}[!t]
\centering%
\includegraphics[width=0.9\textwidth,bb=-246 140 880 650,clip=]{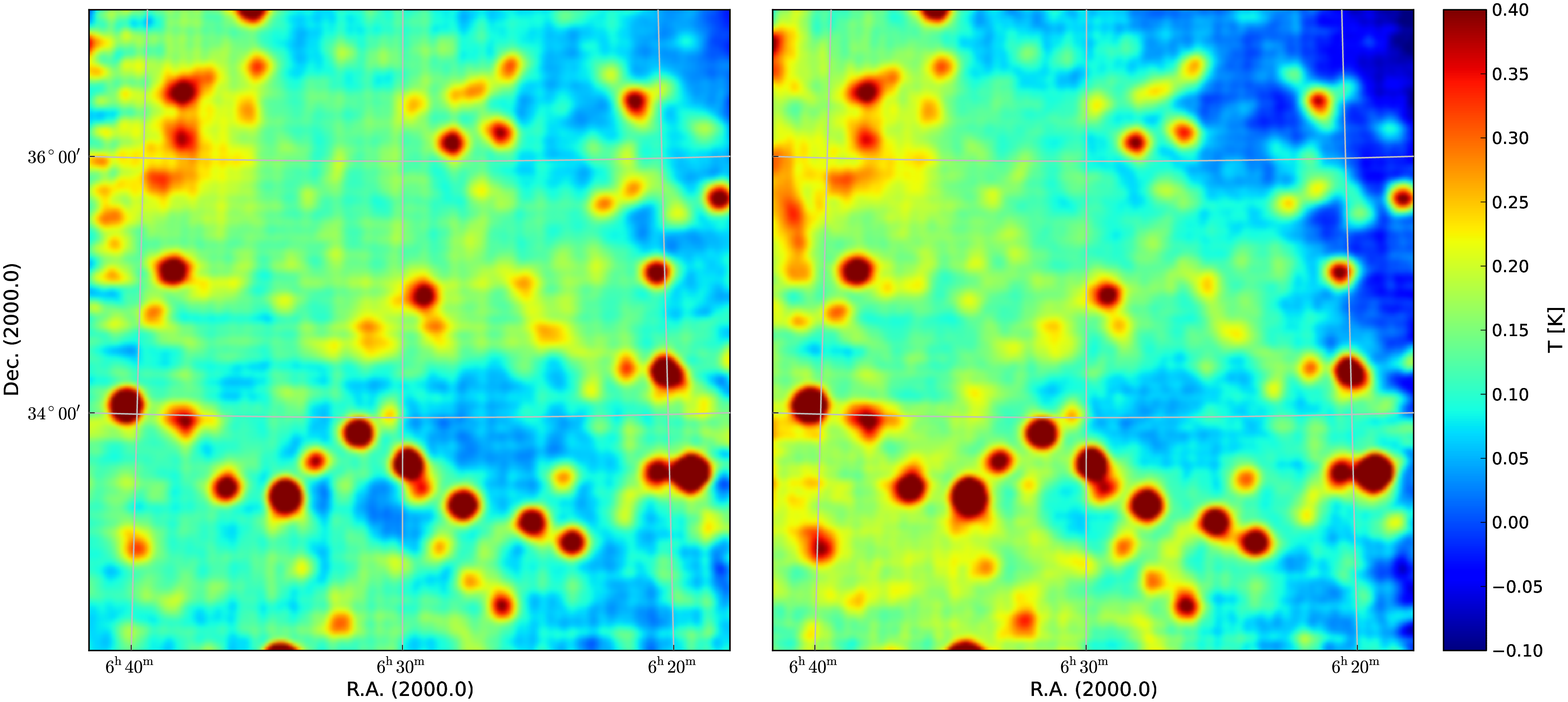}
\caption{Example continuum maps from the Effelsberg--Bonn \ion{H}{i} Survey. Left panel: without basket-weaving. Right panel: with basket-weaving applied. A percentile filter was used to remove elevation-dependent contributions to the system temperature.}%
\label{fig:ebhiscontelevationcontrolled}%
\end{figure*}

After removing elevation-dependent contributions to the baseline level, the data were gridded; see Fig.\,\ref{fig:ebhiscontelevationcontrolled} left panel. The map is already quite acceptable but it can be even more improved with the basket-weaving method as shown in Fig.\,\ref{fig:ebhiscontelevationcontrolled} right panel. Especially, the scan pattern in the upper left quadrant of the image has been substantially suppressed.

\section{Summary}\label{sec:summary}
We have presented a novel technique to solve the basket-weaving problem. Although only offsets and not gains can be solved for (since it is based on linear least-squares), it has several advantages compared to previous approaches:

\begin{itemize}
\item The new algorithm works on gridded data. As a main consequence it is fast (especially if strongly oversampled) and it can be efficiently applied to spectral data. Furthermore, arbitrary scan geometries can be worked with, while other methods rely on rectangular grids \citep[e.g.,][]{sofue79,emerson88}. 
\item Arbitrary basis functions can be used for the offsets in each scan line. In this paper we showed how to apply polynomials of a certain degree, but one could as well work with Fourier series or Legendre polynomials. These basis function can be arbitrarily parametrized. For example it would make sense to use elevation as parameter for more complicated scan geometries since atmospheric effects as one of the main contributors to scan-line artifacts in the data are mainly a function of elevation.
\item It is very easy to handle bad data (e.g., RFI) by masking bad pixels in the maps and associated rows in the basket-weaving equation.
\item Often it is desirable to incorporate prior data, e.g., (lower-resolution) maps from existing surveys or from different wavelengths. Our technique can be easily extended to constrain the solution to be consistent with such priors.
\end{itemize}

\begin{acknowledgements}
We would like to thank Peter M\"{u}ller for carefully reading the manuscript and for useful discussions. Furthermore, we are thankful to Andreas Brunthaler and Carlos Carrasco Gonzalez for providing the 6-cm data and collaboration. Our results are based on observations with the 100-m telescope of the MPIfR (Max-Planck-Institut f\"ur Radioastronomie) at Effelsberg. The research leading to these results has received funding from the European Commission Seventh Framework Programme (FP/2007-2013) under grant agreement No. 283393 (RadioNet3). CCG has been supported by the ERC Advanced Grant GLOSTAR under grant agreement no. 247078. 
\end{acknowledgements}

\bibliographystyle{aa}
\bibliography{references}

\end{document}